\documentclass[aps,preprint,amsmath]{revtex4-1}
\usepackage{graphics}
\begin{document}
\newenvironment{tab}[1]
{\begin{tabular}{|#1|}\hline}
{\hline\end{tabular}}

\title {Elastic vs brittle behaviour of carbon nanotubes, investigated 
through molecular dynamics simulations}

\author{N. Stefanakis}
\affiliation{GMA Gamma measurements and analyses e.K., PO Box 1611, D-72706 Reutlingen, Germany
}  
\date{\today}

\begin{abstract}
The objective of the present paper 
is to investigate the mechanical properties of
carbon nanotubes. We use classical molecular dynamics simulation 
in order to study 
the effect 
of external compression, bending and torsion 
to nanotubes. We find that in the low temperature limit, 
the nanotubes are resilient, 
sustaining extreme strain without signs of brittleness or 
plasticity. For high temperatures and under tension the nanotubes 
show brittle or plastic behavior. 
\end{abstract}
\pacs{}
\maketitle

\section{Introduction}
The carbon nanotubes are formed when the hexagonal graphite sheet 
is rolled to form a seamless cylinder. 
They have important electronic and mechanical 
properties \cite{iijima}. The introduction of pentagons and 
heptagons to the lattice is followed with a positive or 
negative curvature and changes the nanotube shape. 
For their experimental use e.g. as nanometer-scale electronic devices,
wires or as high-strength, light weight materials, it is necessary to know 
the limits of external strain that the nanotubes can sustain.

The mechanical transformation of carbon nanotubes can have 
an effect into the electronic properties of nanotubes only 
when they result to topological changes in the hexagonal 
carbon lattice \cite{stefan,stefan1}.
Recent experiments in transmission electron microscopy of nanotubes
confirm that by elastic deformations of nanotubes no atomic defects are 
formed \cite{chopra}.
 
In this paper we study within classical molecular 
dynamics simulations 
the mechanical properties of carbon nanotubes under external mechanical load. 
We 
examine the effect of mechanical load:
axial compression, bending, and torsion for nanotubes
\cite{yakobson,nardelli}. 

We find that at the low temperature limit, the nanotubes can
sustain extreme strain without signs of brittleness or 
plasticity. However under extreme temperature and appropriate 
strain conditions the nanotubes show effects of plasticity 
or brittleness. At certain levels of strain abrupt changes in tube morphology 
occur which are followed by discontinuities in the elastic energy of the tube.

In the following we describe the methods in Sec II.
In Sec III we present the results for the elastic limit. 
The plastic behaviour is presented in Sec. IV. 
In the last section we present the conclusions.

\section{method}

We used classical molecular dynamics simulations in order to 
study several mechanical transformations of nanotubes. The 
atomic interaction was modeled by the Tersoff-Brenner potential which 
reproduces the lattice constants, binding energies
and elastic constants of graphite and diamond \cite{brenner}. Within this method 
the end atoms were shifted by small steps and remained 
constrained while the rest of the system was relaxed by a conjugate-gradient 
method.

\section{elastic behavior}
\subsection{axial compression}
\begin{figure}
\scalebox{0.3}{\includegraphics{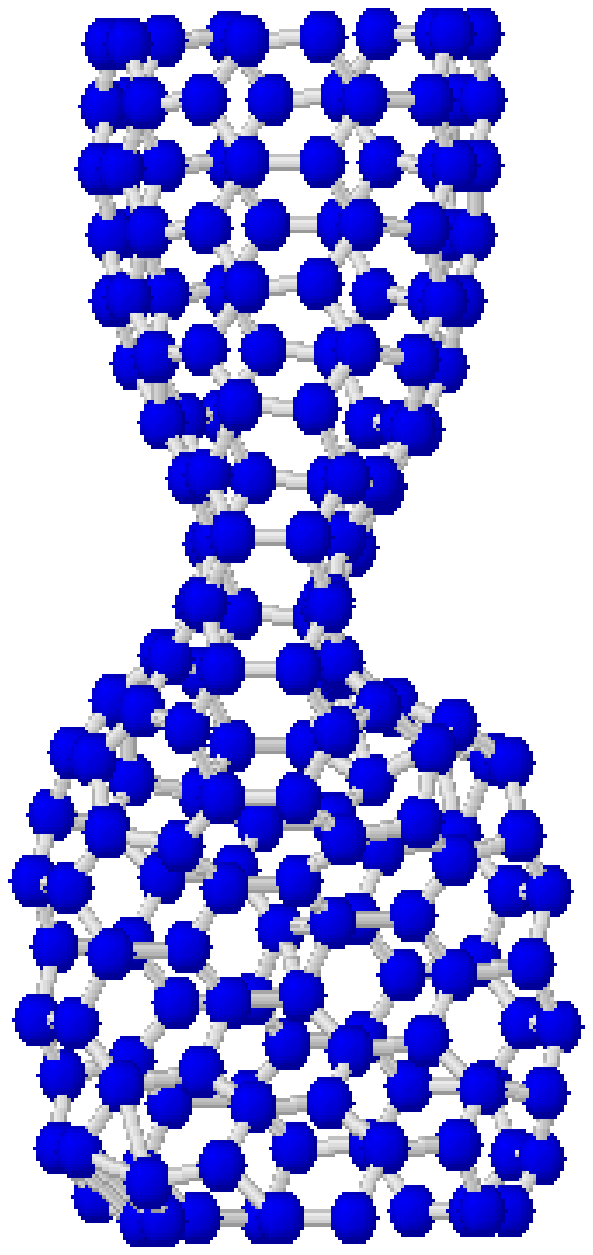}
\includegraphics{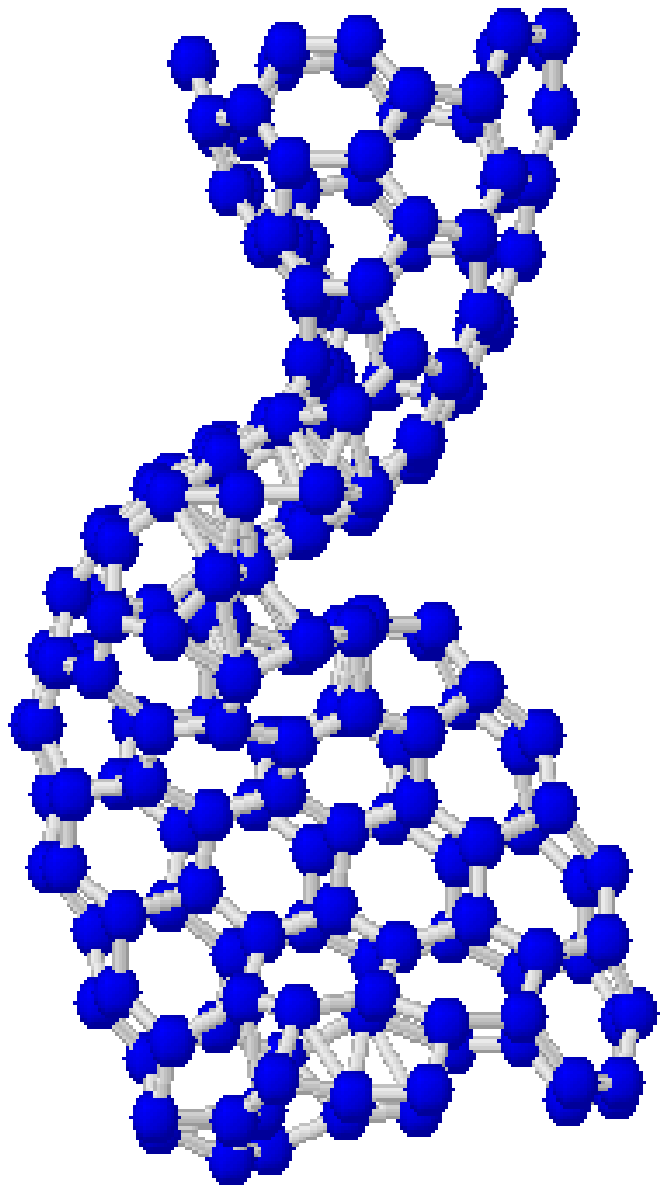}
\includegraphics{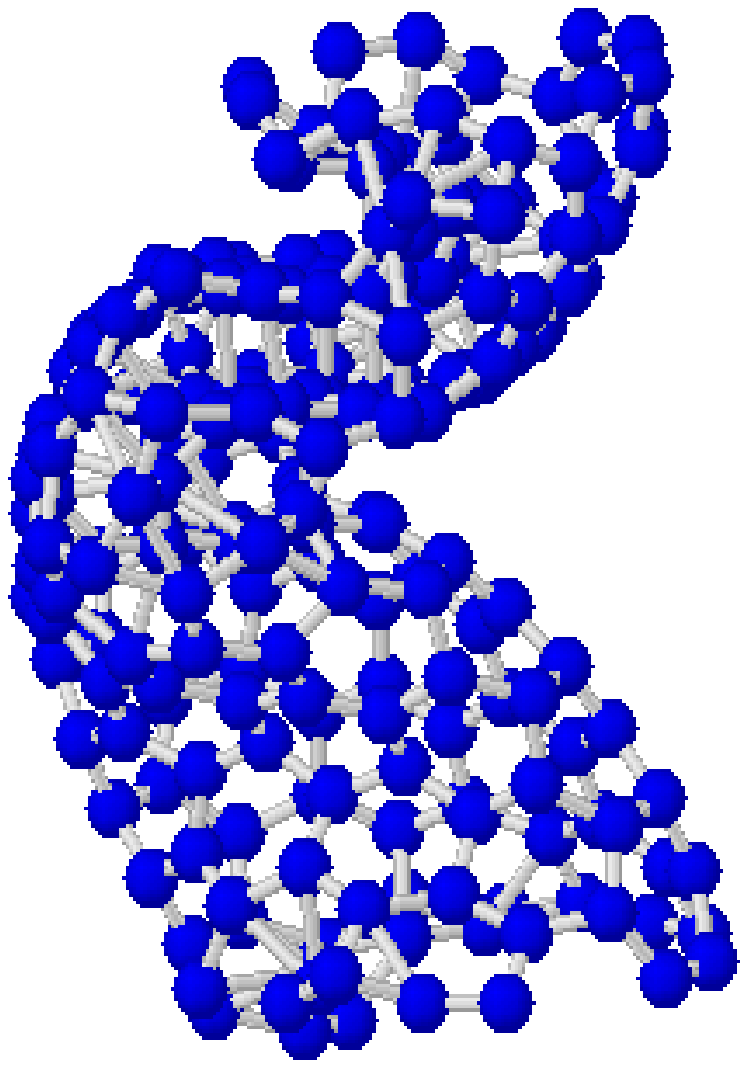}}
\caption{MD simulated armchair $(5,5)$ nanotube of length $21$ layers under axial compression.}
\label{compresssnapshot.fig}
\end{figure}

\begin{figure}
\newpage
\centering\scalebox{0.6}{\includegraphics{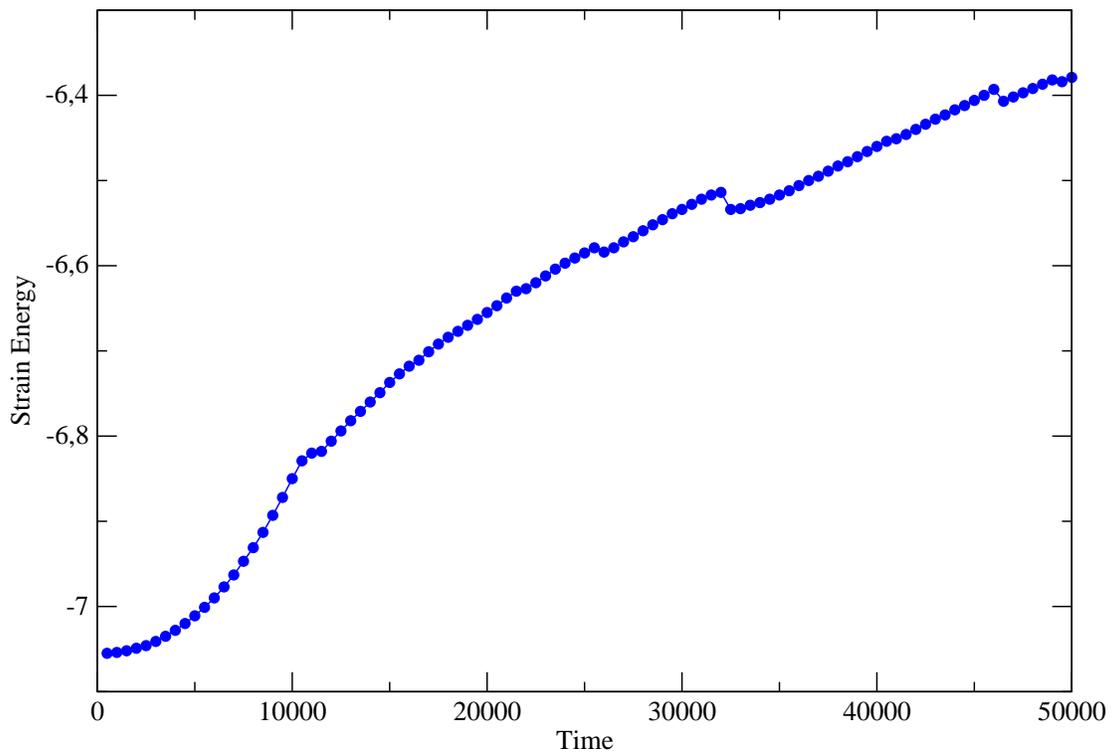}}
\caption{The strain energy of an armchair $(5,5)$ nanotube of length $21$ layers unter axial compression, as a function of time in fs.} 
\label{compress.fig}
\end{figure}

We present in Fig. \ref{compresssnapshot.fig} 
an armchair $(5,5)$ nanotube subject to axial 
compression for different values of the external stress. 
The end atoms are shifted by small steps, 
and are kept at fixed positions in space while the rest 
atoms participate to the molecular dynamics \cite{yakobson}.
At small strains the energy grows quadratic in strain 
(see Fig. \ref{compress.fig}). 
At low values of strain the simulation results can be compared 
with linear analysis results \cite{landau}.
Here the critical value of the strain at which the 
morphological instability occurs, 
depends on the nanotube length.
At higher strains there are 
singularities in the energy versus strain which corresponds 
to structural changes of the system. 
We observed that at the first singulariry the tube buckles to 
a pattern showing two flattenings perpendicular to each other 
(see Fig. \ref{compresssnapshot.fig} left). 
At the next critical value of the strain the tube switches to 
a pattern consisting of three flattenings perpendicular to each 
other still possessing a straight axis. At the next 
critical value of the strain the tube bulks sidewards and only 
a plane of symmetry is preserved 
(see Fig. \ref{compresssnapshot.fig} middle). Finally at higher 
strain values the tubes loses its symmetry  
(see Fig. \ref{compresssnapshot.fig} right).

\subsection{bendig}
\begin{figure}
\scalebox{0.4}{\includegraphics{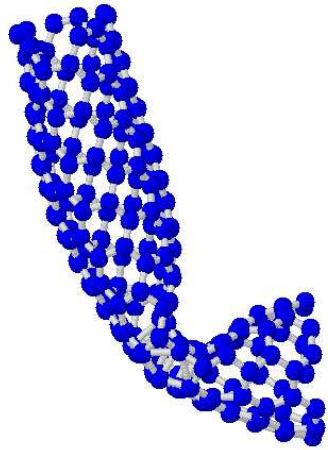}}
\caption{MD simulated armchair $(5,5)$ nanotube of length $21$ layers under bending, after $50000$ fs.
}
\label{bendsnapshot.fig}
\end{figure}

\begin{figure}
\scalebox{0.6}{\includegraphics{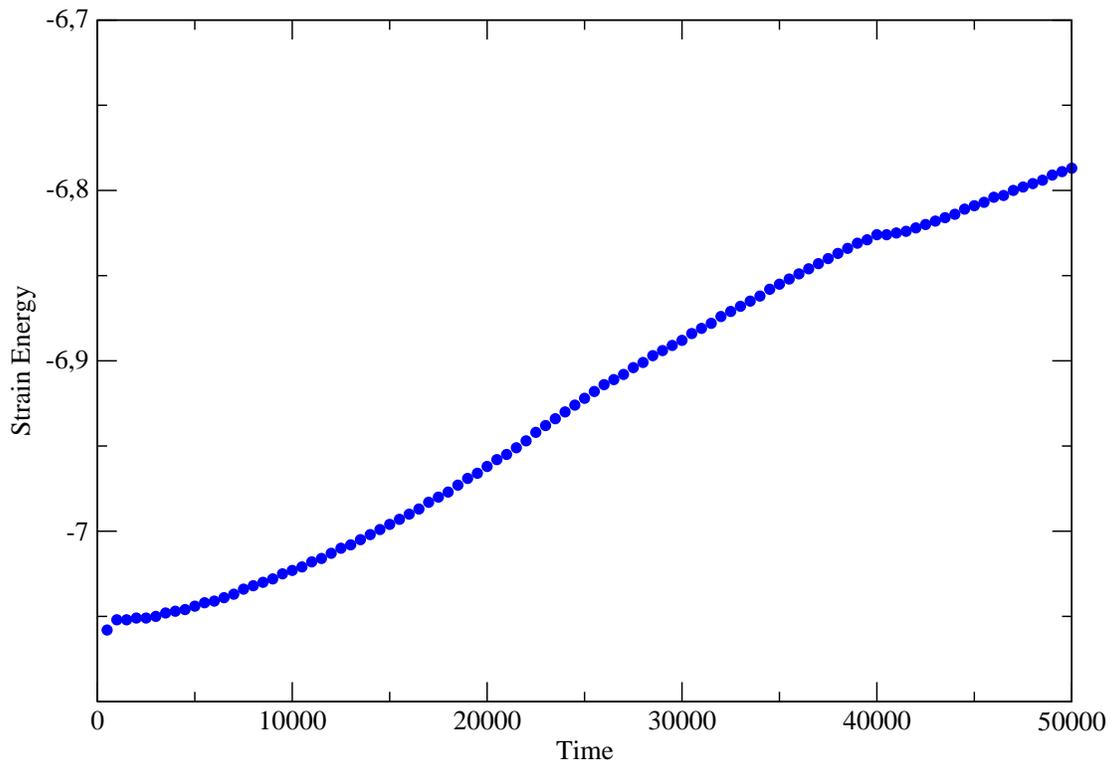}}
\caption{The strain energy of an armchair $(5,5)$ nanotube of length $21$ layers unter bending, as a function of time in fs.} 
\label{bend.fig}
\end{figure}

We present in Fig. \ref{bendsnapshot.fig} an armchair nanotube 
subject to bending load. In simulations a torque is applied and the 
bending angle increases stepwise. At a critical value of the 
bending angle a singularity appears in the energy which corresponds 
to a buckling of the tube where a kink develops (see Fig. \ref{bend.fig}). 
At the 
buckling point, the energy curve 
switches from harmonic to linear.

\subsection{torsion}
\begin{figure}
\centering\scalebox{0.4}{\includegraphics{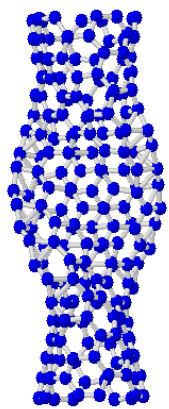}}
\caption{MD simulated armchair $(5,5)$ nanotube of length $21$ layers under torsion after $100000$ fs.}
\label{torsionsnapshot.fig}
\end{figure}

\begin{figure}
\centering\scalebox{0.6}{\includegraphics{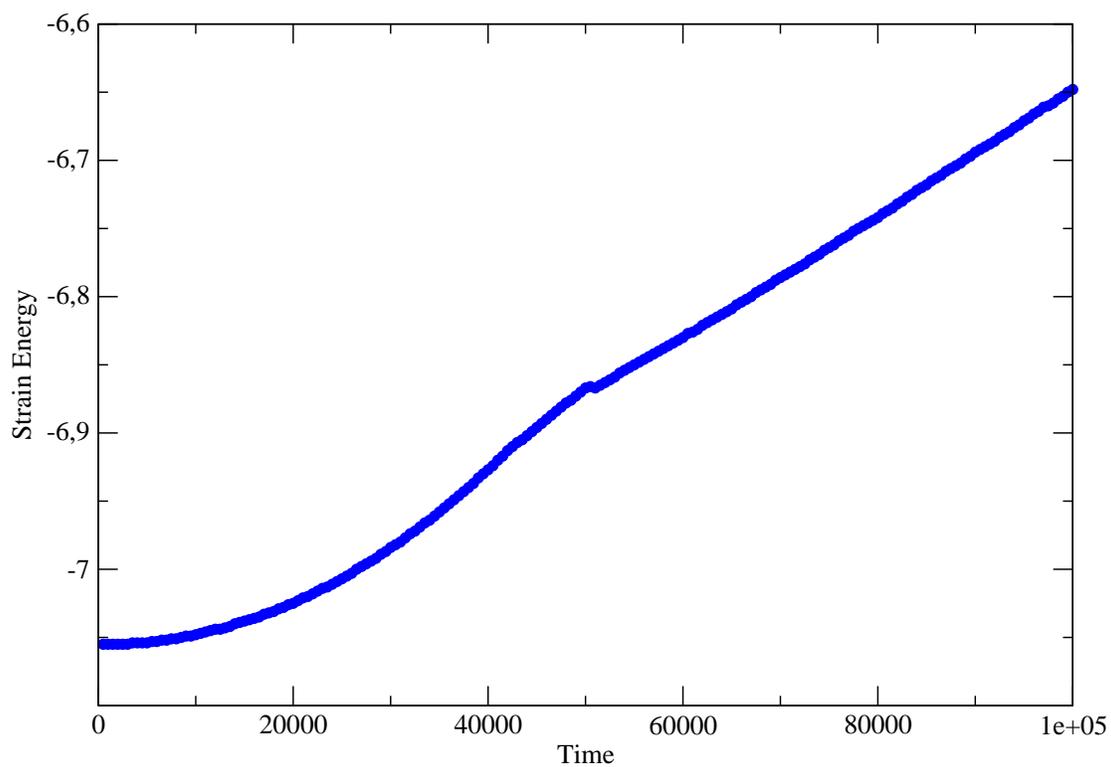}}
\caption{The strain energy of an armchair $(5,5)$ nanotube of length $21$ layers unter torsion, as a function of time in fs.} 
\label{torsion.fig}
\end{figure}

In Fig. \ref{torsionsnapshot.fig} we show an armchair nanotube subject 
to torsion. In simulations the azimuthal angle between the tube 
ends changes. At a critical value of the angle a structural 
phase transformation occurs in which the tube flattens into a
straight axis spiral (see Fig. \ref{torsion.fig}). 

These types of deformations of nanotubes are not combined 
with a change in the topology i.e. the atomic arrangement 
of the hexagonal lattice and demonstrate
that nanotubes show elastic and not brittle behavior even 
at large strengths. 

\section{plastic and brittle behavior}

\begin{figure}
\centering\scalebox{0.6}{\includegraphics{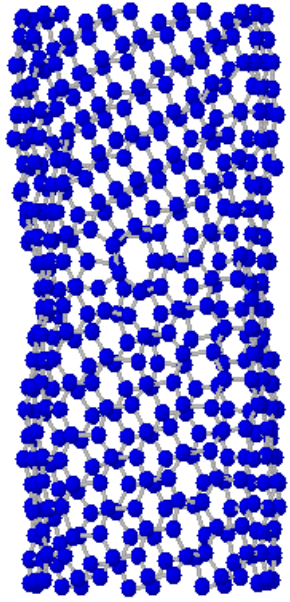}}
\caption{The armchair $(10,10)$ nanotube of length $l=24$. After axial compression of the nanotube at strain $10$ present and temperature $4000 K$, a $5775$ defect appears.
}
\label{4000snapshot.fig}
\end{figure}

We examine in this section the conditions for the formation of 
topological defects in nanotubes. We show that under appropriate 
strain and temperature conditions topological defects 
can be formed which can lead to either ductile or brittle behavior. 
A defect which appears often 
is the bond rotation defect or $5775$ defect which 
is formed from a rotation of a bond for $90$ degrees. 
It is found that this defect is energetically favored for high temperature 
and stress which does not exceed the value of $6\%$. 
We present in Fig. \ref{4000snapshot.fig} an armchair $(10,10)$ 
nanotube at high temperature and axial stress of $10\%$. 
Under these conditions a $5775$ defect appears. 

By reducing the strain and increasing the temperature 
ductile behavior can be observed in which the dislocation pair 
splits to two oppositely oriented dislocations which move. This effect 
can be used in order to change the chirality of the tube. 
On the other hand by increasing the strain under lower temperature 
octagon defects are formed and this leads to brittle behavior \cite{nardelli}. 

\section{conclusions}
We investigated the mechanical properties of
carbon nanotubes. We used classical molecular dynamics simulation 
in order to study 
the effect 
of the external compression, bending and torsion 
to nanotubes. We found that the nanotubes are resilient, 
sustaining extreme strain without signs of brittleness or 
plasticity. 

We also studied non elastic behavior of carbon nanotubes.
We found that under conditions of 
high temperature and small strain, topological defects 
can be formed i.e. bond rotation defects which depending on 
the temperature and strain conditions can evolve to either 
brittle or plastic behavior.

\section{acknowledgments}
Part of this work was done at the 
Center for Nonlinear Phenomena and Complex Systems, 
Free University of Brussels, Code Postal 231, Boulevard du Triomphe,
B-1050, Brussels, Belgium.

This work was supported in part from a stipend from ULB through the 
European research training network HPRN-CT-2002-00198 (DEFINO).

\bibliographystyle{prsty}

\begin{thebibliography}{99}

\bibitem{iijima} S. Iijima,
Nature {\bf 354}, 56 (1991).
\bibitem{stefan} N. Stefanakis,
Phys. Rev. B {\bf 70}, 012502 (2004).
\bibitem{stefan1} N. Stefanakis,
Physics/0502032 (2005).
\bibitem{chopra} N. Chopra, L. Benedict, V. Crespi, M. Cohen, S. Louie and A. Zettl, Nature (London) {\bf 377}, 135 (1995).
\bibitem{yakobson} B.I. Yakobson, C.J. Brabec, and J. Bernholc
Phys. Rev. Lett. {\bf 76}, 2511 (1996).
\bibitem{nardelli} M.B. Nardelli, B.I. Yakobson, and J. Bernholc, 
Phys. Rev. Lett. {\bf 81}, 4656 (1998).
\bibitem{brenner} D.W. Brenner,
Phys. Rev. B {\bf 42}, 9458 (1990).
\bibitem{landau} L.D. Landau and E.M. Lifshitz,
Elasticity Theory, Pergamon, Oxford, 1986.
\end{thebibliography}


\newpage

\vspace{5cm}

\newpage
\vspace{5cm}

\newpage

\newpage
\vspace{5cm}

\newpage

\newpage
\end{document}